\documentclass[a4paper]{article}

\usepackage{INTERSPEECH_v2}
\usepackage{url}

\title{Collaborative Learning for Language and Speaker Recognition}
\name{Lantian Li$^1$, Zhiyuan Tang$^1$, Dong Wang$^{1*}$, Andrew Abel$^2$, Yang Feng$^1$, Shiyue Zhang$^1$}
\address{
  $^1$Center for Speech and Language Technologies, Tsinghua University, China\\
  $^2$Xi'An Jiaotong Liverpool-University, Suzhou, China}
\email{\{lilt,tangzy,fengyang,zhangsy\}@cslt.riit.tsinghua.edu.cn; wangdong99@mails.tsinghua.edu.cn;\\
andrew.abel@xjtlu.edu.cn}

\begin{document}

\maketitle
\begin{abstract}

  This paper presents a unified model to perform language and speaker recognition simultaneously and together.  This model is based on a multi-task recurrent neural network, where the output of one task is fed in as the input of the other, leading  to a collaborative learning framework that can improve both language and speaker recognition by sharing information between the tasks. The preliminary experiments presented in this paper demonstrate that the multi-task model outperforms similar task-specific models on both language and speaker tasks. The language recognition improvement is especially remarkable, which we believe is due to the speaker normalization effect caused by using the information from the speaker recognition component.


\end{abstract}

\section{Introduction}
\label{sec:intro}

	Language recognition (LRE)~\cite{jiri2001lre} and speaker recognition (SRE)~\cite{bimbot2004tutorial} are two important tasks in speech processing.  Traditionally, the research in these two fields seldom acknowledges the other domain, although some there are a number of shared techniques, such as SVM~\cite{campbell2006support}, the i-vector model~\cite{Najim,lei2014novel,Najim2011lang,martinez2011language}, and deep neural models~\cite{ehsan14,heigold2015end,snyderdeep16,lopez2014automatic,lozano2015end,garcia2016stacked,jin2016lid,zazo2016language,kotov2016language}. This lack of overlap can be largely attributed to the intuition that speaker characteristics are language independent in SRE, and dealing with speaker variation is regarded as a basic request in LRE.  This independent processing of language identities and speaker traits, however, is not the way we human beings process speech signals: it is easy to imagine that our brain recognizes speaker traits and language identities simultaneously, and that the success of identifying languages helps discriminate between speakers, and vice versa.

    A number of researchers have noticed that language and speaker are two correlated factors.  In speaker recognition, it has been confirmed that language mismatch indeed leads to serious performance degradation for speaker recognition~\cite{ma2004,Auckenthaler01,Abhinav}, and some language-aware models have been demonstrated successfully~\cite{rozilanguage16}.  In language recognition, speaker variation is seen as a major corruption and is often normalized in the front-end, e.g., by VTLN~\cite{matejka2006brno,gelly2016divide} or CMLLR~\cite{shen2008improved}. These previous studies suggest that speaker and language are inter-correlated factors and should be modelled in an integrated way.

\begin{figure}[htb]
    \centering
    \includegraphics[width=0.8\linewidth]{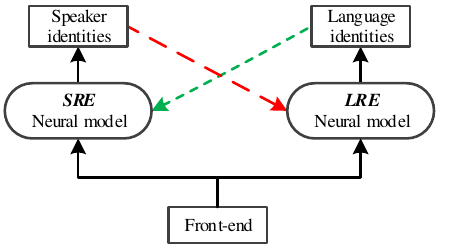}
    \caption{Multi-task recurrent model for language and speaker recognition.}
    \label{fig:diagram}
    \vspace{-1mm}
    \end{figure}

    This paper presents a novel collaborative learning approach which models speaker and language variations in a single neural model architecture. The key idea is to propagate the output of one task to the input of the other, resulting in a multi-task recurrent model. In this way, the two tasks can be learned and inferred simultaneously and collaboratively, as illustrated in Figure~\ref{fig:diagram}.  It should be noted that collaborative learning is a general framework and the component for each task can be implemented using any model, but in this paper, we have chosen to make use of recurrent neural networks (RNN) due to their great potential and good results in various speech processing tasks, including SRE~\cite{heigold2015end,zhang2017end} and LRE~\cite{zazo2016language,gelly2016divide,gonzalez2014automatic,salamea2016use}. Our experiments on the WSJ English database and a Chinese database of a comparable volume demonstrate that the collaborative training method can improve performance on both tasks, and the performance gains on language recognition are especially remarkable.

    In summary, the contributions of this paper are: firstly, we demonstrate that SRE and LRE can be jointly learned by collaborative learning, and that the collaboration benefits both tasks; secondly, we show that the collaborative learning is especially beneficial for language recognition, which is likely to be due to the normalization effect of using the speaker information provided from the speaker recognition component.

   The rest of the paper is organized as follows: we first discuss some related work in Section~\ref{sec:rel}, and then present the collaborative learning architecture in Section~\ref{sec:model}.  The experiments are reported in Section~\ref{sec:exp}, and the paper is concluded in Section~\ref{sec:con}.


\section{Related work}
\label{sec:rel}


	This collaborative learning approach was proposed by Tang et al. for addressing the close relationship between speech and speaker recognition~\cite{tang2016multi}. The idea of multi-task learning for speech signals has been extensively studied, e.g.,~\cite{li2015modeling,qian2016neural}, and more research on this multi-task learning can be found in~\cite{wang2015transfer}. The key difference between collaborative learning and traditional multi-task learning is that the inter-task knowledge share is on-line, i.e., results of one task will impact other tasks, and this impact will be propagated back to itself by the feedback connection, leading to a collaborative and integrated information processing framework.

    The close correlation between speaker traits and language identities is well known to both SRE and LRE researchers. In language recognition, the conventional phonetic approach~\cite{lamel1994language,zissman1996comparison} relies on the compositional speech recognition system to deal with the speaker
    variation. In the HMM-GMM era, this often relied on various front-end normalization techniques, such as vocal track length normalization (VTLN)~\cite{matejka2006brno,gelly2016divide} and constrained maximum likelihood linear regression (CMLLR)~\cite{shen2008improved}.  In the HMM-DNN era, a DNN model has the natural capability to normalize speaker variation when sufficient training data is available. This capability has been naturally used in i-vector based LRE approaches~\cite{song2013vector,tian2016investigation}.  However, for pure acoustic-based DNN/RNN methods, e.g.,~\cite{jin2016lid,zazo2016language}, there is limited research into speaker-aware learning for LRE.

    For speaker recognition, language is often not a major concern, perhaps due to the a widely held assumption that speaker traits are language independent. However from the engineering perspective, language mismatch has been found to pose a serious problem due to the different patterns of acoustic space in different languages, according to their own phonetic systems~\cite{ma2004,Auckenthaler01,Abhinav}.  A simple approach is to train a multi-lingual speaker model by data pooling~\cite{ma2004,Auckenthaler01},
    but this approach does not model the correlation between language identities and speaker traits.  Another potential approach is to treat language and speaker as two random variables and represent them by a linear Gaussian model~\cite{lu2009jfa}, but this linear Gaussian assumption is perhaps too strong.

    The collaborative learning approach benefits both tasks. For SRE, the language information provided by LRE helps to identify acoustic units that the recognition should focus on, and for LRE, the speaker information provided by SRE helps to normalize the speaker variation. It is important to note that the models for these two tasks are jointly optimized, and that the information from both tasks during decoding. This means that the collaborative learning is collaborative in both model training and inference.








\section{Multi-task RNN and collaborative learning}
  \label{sec:model}

  This section first presents the neural model structure for single tasks, and then extends this to the multi-task recurrent model for collaborative learning.

  \subsection{Basic single-task model}

  For the work in this paper we have chosen a particular RNN, the long short-term memory (LSTM)~\cite{Sepp1997lstm} approach to build the baseline single-task systems for SRE and LRE.  LSTM has been shown to deliver good performance for both SRE~\cite{heigold2015end} and LRE~\cite{zazo2016language,gelly2016divide,gonzalez2014automatic}.  In particular, the recurrent LSTM structure proposed in~\cite{sak2014long} is used here, as shown in Figure~\ref{fig:lstm}, and the associated computation is as follows:

    \begin{figure}[htb]
    \centering
    \includegraphics[width=0.95\linewidth]{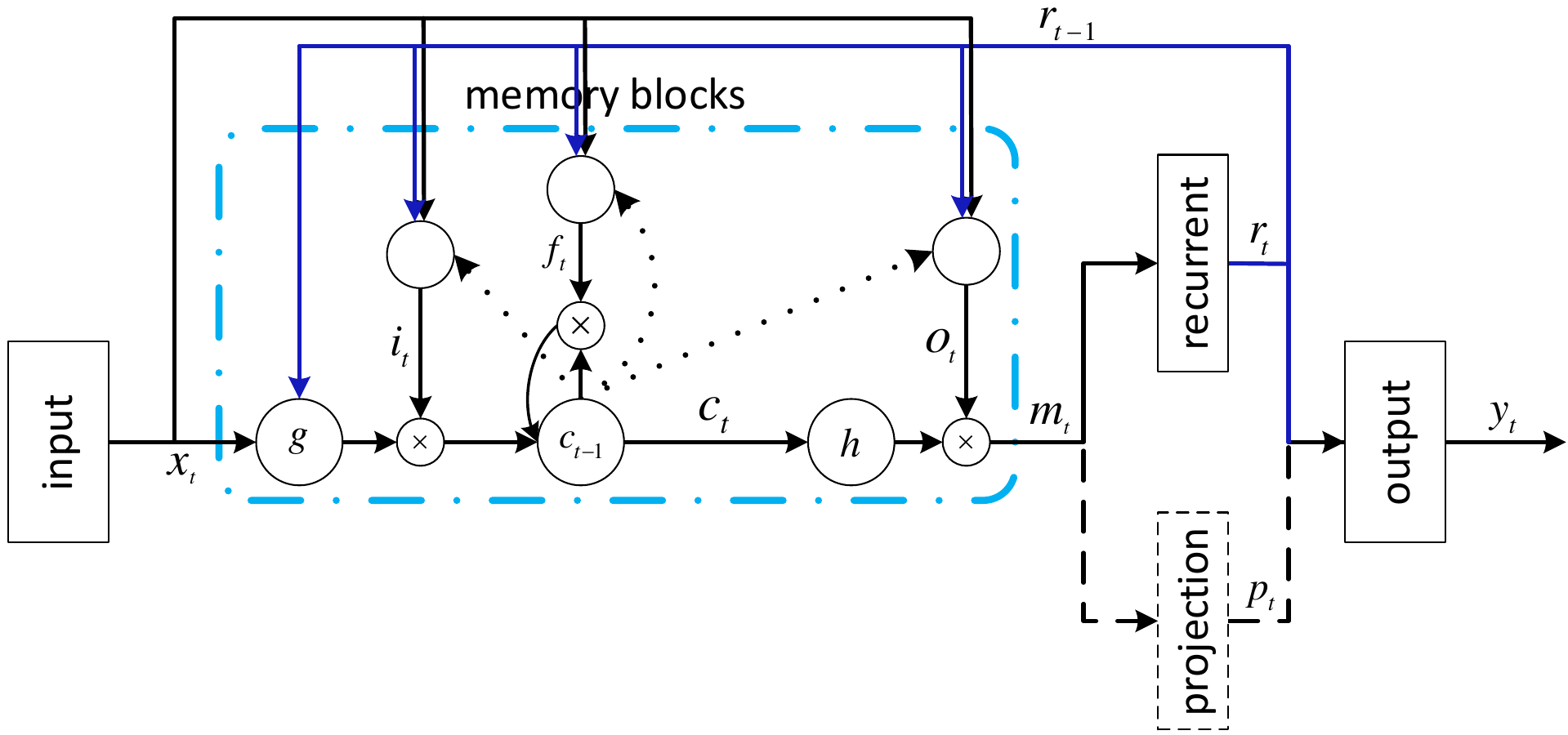}
    \caption{Basic recurrent LSTM model for LRE and SRE single-task baselines.}
    \label{fig:lstm}
      \vspace{-2mm}
    \end{figure}

  \vspace{-2mm}
  \begin{eqnarray}
    i_t &=& \sigma(W_{ix}x_{t} + W_{ir}r_{t-1} + W_{ic}c_{t-1} + b_i) \nonumber\\
    f_t &=& \sigma(W_{fx}x_{t} + W_{fr}r_{t-1} + W_{fc}c_{t-1} + b_f) \nonumber\\
    c_t &=& f_t \odot c_{t-1} + i_t \odot g(W_{cx}x_t + W_{cr}r_{t-1} + b_c) \nonumber\\
    o_t &=& \sigma(W_{ox}x_t + W_{or}r_{t-1} + W_{oc}c_t + b_o) \nonumber\\
    m_t &=& o_t \odot h(c_t) \nonumber\\
    r_t &=& W_{rm} m_t \nonumber\\
    p_t &=& W_{pm} m_t \nonumber\\
    y_t &=& W_{yr}r_t + W_{yp}p_t + b_y \nonumber.
  \end{eqnarray}

  \noindent In the above equations, the $W$ terms denote weight matrices and the $b$ terms denote bias vectors. $x_t$ and $y_t$ are the input and output vectors; $i_t$, $f_t$, $o_t$ represent the input, forget and output gates respectively; $c_t$ is the cell and $m_t$ is the cell output. $r_t$ and $p_t$ are the two output components derived from $m_t$, in which $r_t$ is recurrent and used as an input of the next time step, while $p_t$ is not recurrent and contributes to the present output only.  $\sigma(\cdot)$ is the logistic sigmoid function, and $g(\cdot)$ and $h(\cdot)$ are non-linear activation functions, often chosen to be hyperbolic. $\odot$ denotes the element-wise multiplication.

  \subsection{Multi-task recurrent model}

\begin{figure}[htb]
    \centering
    \includegraphics[width=1\linewidth]{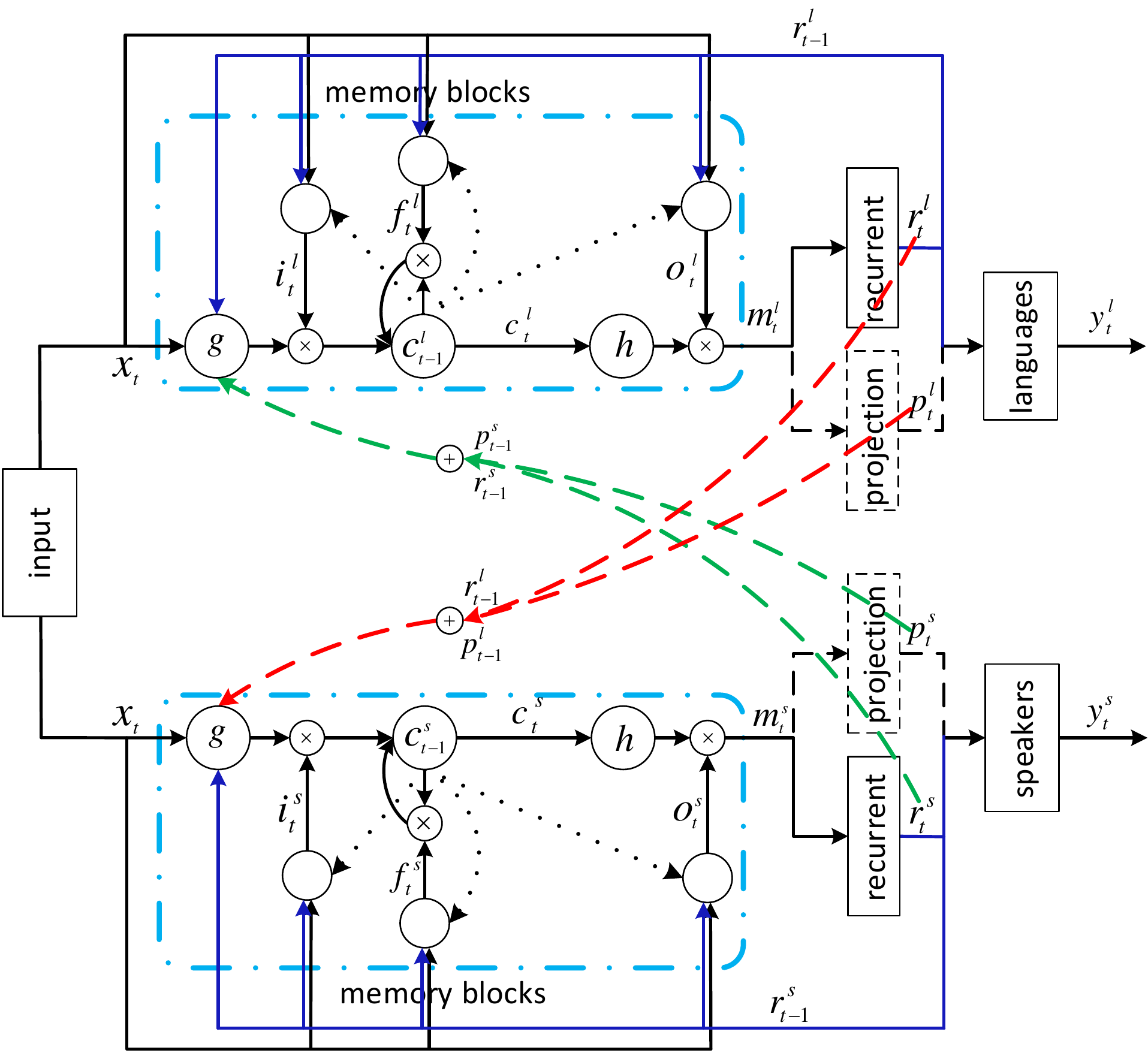}
    \caption{Multi-task recurrent learning for LRE and SRE.}
    \label{fig:multi}
    \end{figure}

  The basic idea of the multi-task recurrent model, as shown in Figure~\ref{fig:diagram}, is to use the output of one task at the current time step as an auxiliary input into the other task at the next step.  In this study, we use the recurrent LSTM model to build the LRE and SRE components, and then combine them with a number of inter-task recurrent connections. This results in a multi-task recurrent model, by which LRE and SRE can be trained and inferred in a collaborative way.  The complete model structure is shown in Figure~\ref{fig:multi}, where the superscripts $l$ and $s$ denote the LRE and SRE task respectively, and the dashed lines represent the inter-task recurrent connections.

  A multitude of possible model configurations can be selected. For example, feedback information can be extracted from the cell $c_t$ or cell output $m_t$, or from the output component $r_t$ or $p_t$; the feedback information can be propagated to the input variable $x_{t}$, the input gate $i_t$, the output gate $o_t$, the forget gate $f_t$, or the non-linear function $g(\cdot)$.

  Given the above alternatives, the multi-task recurrent model is rather flexible.  The structure shown in Figure~\ref{fig:multi} is one simple example, where the feedback information is extracted from both the recurrent projection $r_t$ and the non-recurrent projection $p_t$, and propagated to the non-linear function $g(\cdot)$.  Using the feedback, the computation for LRE is given as follows:

  \begin{eqnarray}
    i^l_t &=& \sigma(W^l_{ix}x_{t} + W^l_{ir}r^l_{t-1} + W^l_{ic}c^l_{t-1} + b^l_i) \nonumber\\
    f^l_t &=& \sigma(W^l_{fx}x_{t} + W^l_{fr}r^l_{t-1} + W^l_{fc}c^l_{t-1} + b^l_f) \nonumber\\
    g^l_t &=&  g(W^l_{cx}x^l_t + W^l_{cr}r^l_{t-1} + b^l_c +    \underline{ W^{ls}_{cr}r^s_{t-1} + W^{ls}_{cp}p^s_{t-1}  } ) \nonumber\\
    c^l_t &=& f^l_t \odot c^l_{t-1} + i^l_t \odot g^l_t \nonumber\\
    o^l_t &=& \sigma(W^l_{ox}x^l_t + W^l_{or}r^l_{t-1} + W^l_{oc}c^l_t + b^l_o) \nonumber\\
    m^l_t &=& o^l_t \odot h(c^l_t) \nonumber\\
    r^l_t &=& W^l_{rm} m^l_t \nonumber\\
    p^l_t &=& W^l_{pm} m^l_t \nonumber\\
    y^l_t &=& W^l_{yr}r^l_t + W^l_{yp}p^l_t + b^l_y \nonumber 
  \end{eqnarray}

  \noindent and the computation for SRE is given as follows:

  \begin{eqnarray}
    i^s_t &=& \sigma(W^s_{ix}x_{t} + W^s_{ir}r^s_{t-1} + W^s_{ic}c^s_{t-1} + b^s_i) \nonumber\\
    f^s_t &=& \sigma(W^s_{fx}x_{t} + W^s_{fr}r^s_{t-1} + W^s_{fc}c^s_{t-1} + b^s_f) \nonumber\\
    g^s_t &=& g(W^s_{cx}x^s_t + W^s_{cr}r^s_{t-1} + b^s_c +    \underline{ W^{sl}_{cr}r^l_{t-1} + W^{sl}_{cp}p^l_{t-1}  } ) \nonumber\\
    c^s_t &=& f^s_t \odot c^s_{t-1} + i^s_t \odot g^s_t \nonumber\\
    o^s_t &=& \sigma(W^s_{ox}x^s_t + W^s_{or}r^s_{t-1} + W^s_{oc}c^s_t + b^s_o) \nonumber\\
    m^s_t &=& o^s_t \odot h(c^s_t) \nonumber\\
    r^s_t &=& W^s_{rm} m^s_t \nonumber\\
    p^s_t &=& W^s_{pm} m^s_t \nonumber\\
    y^s_t &=& W^s_{yr}r^s_t + W^s_{yp}p^s_t + b^s_y \nonumber 
  \end{eqnarray}

  \subsection{Model training}

  The model can be trained `completely', where each training sample is labelled by both speaker and language, or `incompletely' where only one task label is available.  Our previous research has demonstrated that both cases are suitable ~\cite{tang2016multi}. In this preliminary study, we have focused on using `completely' training. The natural stochastic gradient descent (NSGD) algorithm ~\cite{povey2014parallel} is employed to train the model.

  \section{Experiments}
  \label{sec:exp}

  This section first describes the data profile, and presents the baseline systems.  Finally, experimental results of our collaborative learning approach are given.

  \subsection{Data}

  Two databases were used to perform the experiment: the WSJ database in English and the CSLT-C300 database in Chinese\footnote{This database was collected by our institute for commercial usage, so we cannot release the wave data, but the Fbanks and MFCCs in the Kaldi format have been published online. See \url{http://data.cslt.org}. The Kaldi recipe to reproduce the results is also available there.}.  All the utterances in both databases were labelled with both language and speaker identities. The development set involves two subsets: WSJ-E200, which contains $200$ speakers ($24,031$ utterances) selected from WSJ, and CSLT-C200, which contains $200$ speakers ($20,000$ utterances) selected from the CSLT-C300 database. The development set was used to train the i-vector, SVM, and multi-task recurrent models.

  The evaluation set contains an English subset WSJ-E110, which contains $110$ speakers selected from WSJ, and a Chinese subset CSLT-C100, which contains $100$ speakers selected from the CSLT-C300 database. For each speaker in each subset, $10$ utterances were used to enrol its speaker and language identity, and the remaining $13,236$ English utterances and $9,000$ Chinese utterances were used for testing. For SRE, the test is pair-wised, leading to $13,236$ target trials and $1,442,724$ imposter trials in English, plus $9,000$ target trials and $891,000$ Chinese imposter trials.  For LRE, the number of test trials is the same as the number of test utterances, which is $13,236$ for English trials and $9,000$ for Chinese trials.


  \subsection{LRE and SRE baselines}

  Here, we first present the LRE and SRE baselines.  For each task, two baseline systems were constructed, one based on i-vectors (still state-of-the art),
  and the other, based on LSTM.  All experiments were conducted with the Kaldi toolkit~\cite{povey2011kaldi}.

  \subsubsection{i-vector baseline}

  For the i-vector baseline, the acoustic features were $39$-dimensional MFCCs.  The number of Gaussian components of the universal background model (UBM) was $1,024$, and the dimension of the i-vectors was $200$.  The resulting i-vectors were used to conduct both SRE and LRE with different scoring methods.  For SRE, we consider the simple Cosine distance, as well as the popular discriminative models LDA and PLDA; for LRE, we consider Cosine distance and SVM. All the discriminative models were trained on the development set.

  The results of the SRE baseline are reported in Table~\ref{tab:base-sre}, in terms of equal error rate (EER).  We tested two scenarios, one is a Full-length test which uses the entire enrolment and test utterance; the other is a Short-length test which involves only $1$ second of speech (sampled from the original data after voice activity detection is applied). In both scenarios, the language of each test is assumed to be known in advance, i.e., the tests on English and Chinese
  datasets are independent.

      \begin{table}[th]
        \caption{SRE baseline results.}
        \label{tab:base-sre}
        \centering
          \begin{tabular}{|l|l|l|c|c|c|}
            \hline
            Test  &System & Dataset&\multicolumn{3}{|c|}{EER(\%)} \\
            \cline{4-6}
                     &    &       & Cosine   & LDA    & PLDA \\
            \hline
            Full     &i-vector & English         &  0.88    & 0.70   & {\bf 0.62} \\
                     &         & Chinese        &  1.28    & 0.97   & {\bf 0.84} \\
            \cline{2-6}
                     &r-vector & English        &  1.25    & 1.38   & 3.57 \\
                     &         & Chinese          &  1.70    & 1.61   & 4.93 \\
            \hline
            \hline
            Short    &i-vector & English      & 7.00   & 4.01     & 3.47 \\
                     &         & Chinese       & 9.12   & 6.16     & 5.69 \\
            \cline{2-6}
                     &r-vector & English        & 3.27   & {\bf 2.70}     & 7.88 \\
                     &         & Chinese        & 4.77   & {\bf 3.99}     & 8.21 \\
            \hline
          \end{tabular}
      \end{table}

     LRE is an identification task, with the purpose being to discriminate between two languages (English and Chinese). We therefore use identification error rate (IDR)~\cite{yinbo2002lid} to measure performance, which is the fraction of the identification mistakes in the total number of identification trials. For a more thorough comparison, the number of identification errors (IDE) is also reported.  The results of the i-vector/SVM baseline system are reported in Table~\ref{tab:base-lre}.

      \begin{table}[htp]
        \caption{LRE baseline results.}
        \label{tab:base-lre}
        \centering
          \begin{tabular}{|l|l|c|c|}
            \hline
           Test &  System          &   IDR(\%)  & IDE\\
            \hline
           Full & i-vector/Cosine  &  3.43    &   763 \\
                & i-vector/SVM     &  {\bf 0.01}   & 2   \\
                \cline{2-4}
                & r-vector/Cosine  &  0.11    &   25 \\
                & r-vector/SVM     &  0.21    &   47 \\
                & r-vector/Softmax &  0.13    &   29 \\
            \hline
            \hline
           Short & i-vector/Cosine &  10.21   &   2270 \\
                 & i-vector/SVM    &  1.40    &   311 \\
                 \cline{2-4}
                 & r-vector/Cosine &  0.98    &   218 \\
                 & r-vector/SVM    &  0.63    &   139 \\
                 & r-vector/Softmax&  {\bf 0.58}   &  129    \\
            \hline
          \end{tabular}
      \end{table}




\subsubsection{r-vector baseline}

  The r-vector baseline is based on the recurrent LSTM structure shown in Figure~\ref{fig:lstm}.  The SRE and LRE systems use the same configurations: the dimensionality of the cell was set to $1,024$, and the dimensionality of both the recurrent and non-recurrent projections was set to $100$. For the SRE system, the output corresponds to the $400$ speakers in the training set; For LRE, the output corresponds to the two languages to identify.  The output of both projections were concatenated and averaged over all the frames of an utterance, resulting in a $200$-dimensional `r-vector' for that utterance.  The r-vector derived from the SRE system represents speaker characters, and the r-vector derived from the LRE system represents the language identity.

  As in the i-vector baseline, decisions were made based on distance between r-vectors, measured by either the Cosine distance or some discriminative models. The
  same discriminative models as in the i-vector baseline were used, except that in the LRE system, the softmax outputs of the task-specific LSTMs can be directly used to identify language. The results are shown in Table~\ref{tab:base-sre} and Table~\ref{tab:base-lre} for SRE and LRE, respectively.

  The results in Table~\ref{tab:base-sre} show that for SRE, the i-vector system with PLDA performs better than the r-vector system in the Full-length test. However, in the Short-length test, the r-vector system is clearly better. This is understandable as the i-vector model is generative and relies on sufficient data to estimate the data distribution; the LSTM model, in contrast, is discriminative and the speaker information can be extracted with even a single frame. Moreover, the PLDA model works very well for the i-vector system, but rather poor for the r-vector system.  We estimate that this could be due to the unreliable Gaussian assumption for the residual noise by PLDA.  A pair-wised t-test confirms that the performance advantage of the r-vector/LDA system over the i-vector/PLDA system is statistically significant ($p$ $<$ 1e-5).

  The results in Table~\ref{tab:base-lre} show a similar trend, that the i-vector system (with SVM) works well in the full-length test, but in the short-length test, the r-vector system shows much better performance, even with the simple Cosine distance. Again, this can be explained by the fact that the i-vector model is generative, while the r-vector model is discriminative. The advantage of the r-vector model on short utterances has previously been observed, both for LRE~\cite{zazo2016language} and SRE~\cite{snyderdeep16}.


  \subsection{Collaborative learning}

  The multi-task recurrent LSTM system, as shown in Figure~\ref{fig:multi}, was constructed by combining the LRE and SRE r-vector systems, with inter-task recurrent connections augmented.  Following research in~\cite{tang2016multi}, we selected the output of the recurrent projection layer as the feedback information, and tested several configurations, where the feedback information from one task is propagated into different components of the other task.  The results are reported in Tables~\ref{tab:j-sre} and~\ref{tab:j-lre} for SRE and LRE, where $i,f,o$ denotes the input, forget and output gates, and $g$ denotes the non-linear function.

    \begin{table}[htbp]
        \caption{SRE results with collaborative learning.}
        \label{tab:j-sre}
        \vspace{-2mm}
        \centering
          \begin{tabular}{|cccc|c|c||c|c|}
            \hline
            \multicolumn{4}{|c|}{Feedback}              &        \multicolumn{4}{|c|}{EER(\%)}        \\
            \cline{5-8}
            \multicolumn{4}{|c|}{Input}            &\multicolumn{2}{|c||}{Full} & \multicolumn{2}{|c|}{Short}\\
            \hline
              $i$     & $f$     &  $o$   &   $g$        & Eng. &  Chs.  & Eng. &  Chs.   \\
            \hline
            \multicolumn{4}{|c|}{r-vector Baseline}      &  1.38   &  1.61       &  2.70   &  3.99  \\
            \hline
              $\surd$&         &         &              &  1.27   &   1.43      &  2.50   &  3.61  \\
                     & $\surd$ &         &              &  1.38   &   1.38      &  2.55   &  3.52  \\
                     &         & $\surd$ &              &  {\bf 1.19}   &  {\bf 1.31}      &  {\bf 2.48}   &  3.66  \\
                     &         &         & $\surd$      &  1.37   &   1.48      &  2.67   &  {\bf 3.52}   \\
              $\surd$&$\surd$  & $\surd$ & $\surd$      &  1.32   &   1.31      &  2.52   &  3.69  \\
            \hline
          \end{tabular}
      \end{table}

    \begin{table}[htbp]

        \caption{LRE results with collaborative learning.}
        \label{tab:j-lre}
        \vspace{-2mm}
        \centering
        \resizebox{0.47\textwidth}{!}{ %
          \begin{tabular}{|cccc|c|c|c||c|c|c|}
            \hline
            \multicolumn{4}{|c|}{Feedback}              &  \multicolumn{6}{|c|}{IDE } \\
            \cline{5-10}
            \multicolumn{4}{|c|}{Input}                 &  \multicolumn{3}{|c||}{Full}     &  \multicolumn{3}{|c|}{Short}     \\
            \hline
              $i$     & $f$     &  $o$   &   $g$        &  Cosine & SVM & Softmax &   Cosine & SVM & Softmax   \\
            \hline
             \multicolumn{4}{|c|}{r-vector Baseline}     &    25 &  47  &  29   &    218     &  139   &  129    \\
            \hline
              $\surd$&         &         &              &    5  &  2  &   0    &    11      &   6  &   2   \\
                     & $\surd$ &         &              &    1  &  0  &   0    &    3       &   1  &   1   \\
                     &         & $\surd$ &              &    11 &  2  &   0    &    21      &   8  &   3   \\
                     &         &         & $\surd$      &    0  &  0  &   1    &    2       &   2  &   1   \\
              $\surd$&$\surd$  & $\surd$ & $\surd$      &    6  &  2  &   0    &    17      &   10 &   2   \\
            \hline
          \end{tabular}
      }
      \vspace{-1mm}
     \end{table}

  The results show that collaborative learning provides consistent performance improvement on both SRE and LRE, regardless of which component the feedback is applied to.  The results show that the output gate is an appropriate component for SRE to receive  the feedback, whereas for LRE, the forget gate seems a more suitable choice. However, these observations are based on relatively small databases. More experiments on large data are required to confirm and understand these observations. Finally, it should be highlighted that the collaborative training provides very impressive performance gains for LRE: it significantly improves the single-task r-vector baseline, and beats the i-vector baseline even on the full-length task. This is likely to be because the LRE model trained with the limited training data is largely disturbed by the speaker variation, and the language information provided by the SRE system plays a valuable role of speaker normalization.


\section{Conclusions}
  \label{sec:con}

 This paper proposed a novel collaborative learning architecture that performs speaker and language recognition as a single and unified model, based on a multi-task recurrent neural network. These preliminary experiments demonstrated that the proposed approach can deliver consistent performance improvement over the single-task baselines for both SRE and LRE. The performance gain on LRE is particularly impressive, which we suggest could be due to the effect of speaker normalization.  Future work involves experimenting with large databases and analyzing the properties of the collaborative mechanism, e.g., trainability, stability and extensibility.



\bibliographystyle{IEEEtran}
{\footnotesize
\bibliography{refs}
}

\end{document}